\documentclass[12pt]{iopart}

\usepackage{iopams}
\usepackage{graphicx}
\usepackage{amssymb}
\usepackage{amsfonts}
\usepackage{amsbsy}

\renewcommand{\vec}[1]{\mbox{\boldmath$\mathrm{#1}$}}

\begin{document}

\title[Thermodynamics of ME chain]{Electric field-effect on thermodynamics of multiferroic spin chain}
\author{Chenglong Jia and Jamal Berakdar}
\address{Institut f\"ur Physik, Martin-Luther Universit\"at Halle-Wittenberg,06120 Halle (Saale), Germany}

\begin{abstract}
Thermodynamic properties of the helical spins with magnetoelectric coupling is simulated by Monte Carlo algorithm.  
It is shown that the spin-spin correlations are affected by the thermal fluctuations and the electric field. Results for the temperature-dependence of ferroelectric polarization and the susceptibility are also presented in connection with the multiferroic phase transition.
\end{abstract}

%\pacs{}

\maketitle

\section{Introduction}
The efficient control of magnetism in terms of the electric field in the solid has been highly desired for many years due to their potential application and fundamental curiosity.  By now the most straightforward way is to enhance the magnetoelectric (ME) coupling in the multiferroics, i.e., materials with both ferroelectric (FE) and magnetic orders \cite{ME}. Here the ferroelectricity results from the loss of inversion symmetry induced by the spatial variation of the spin order. Recent experiments indicate that both collinear and noncolinear magnetic order can have a potential to produced a spontaneous FE polarization $\vec P$ \cite{Multiferroics}.  Among them, the multiferroics with spiral spins are particularly interesting \cite{RMnO3}. Every nearest-neighbor spin pair produces a local $\vec P_i$, which can be well described by the spin-current model \cite{KNB}, $$\vec P_i \sim e_{ij} \times (S_i \times S_j)$$ with $e_{ij}$ being the unit vector connecting two neighboring spins $S_i$ and $S_j$. Since the FE behavior is highly sensitive to the applied electric field, the magnetic action is then converted to the electric response.  The static ME effect has been successfully investigated experimentally \cite{E-Control} and theoretically \cite{JB} for the helimagnet oxides, which would be expected to bring about entirely new concepts in the design of next generation devices. On the other hand, the dynamical exchange-striction induces a biquadratic interaction between the spins and transverse phonons  \cite{d-ME} resulting in the hybridization between the spin wave and the fluctuation of the electric polarization, which leads to a new collective ME excited mode, so-called electromagnon,  providing another promising way to enhance the strength of ME coupling by using the resonance in the dynamical regime \cite{electromagnon,d-KNB}.  In this paper,  the spin-current model is revisited by using the Monte-Carlo simulation with classical Heisenberg spins \cite{MC-Book}.  We show how the spin-spin correlations are affected by the thermal fluctuations and the electric field. It is confirmed that the long-range multiferroic order is stabilized by applying electric field that leads to a FE phase with spontaneously broken spin chiral symmetry.  The polarization susceptibility shows an electric field-dependent phase transition. 

\section{Spin current model}
We start with a one-dimensional spin chain along the $x$-axis with a frustrated spin interaction. Taking account of electric energy, an effective $J_1-J_2$ model with the ME coupling is introduced as follows,
\begin{eqnarray}
&& \hat{H} = - J_1 \sum_{\langle ij \rangle _{nn}} S_i \cdot S_j +J_2 \sum_{\langle lm \rangle _{nnn}} S_l \cdot S_m - \vec{E} \cdot \vec {P}, \\
&& \vec{P} = \frac{1}{N} \sum_i [\hat{e}_{x} \times (S_i \times S_{i+1})] 
\end{eqnarray}
where $\vec{E} = (0,0, E)$ is the external electric field. The notation $\langle ij \rangle_{nn}$ indicates  nearest-neighboring (nn)  $i$ and $j$, and $\langle lm \rangle_{nnn} $ corresponds the next-nearest-neighboring (nnn) $l$ and $m$. The competition between the nn ferromagnetic interaction ($- J_1 < 0$) and the nnn antiferromagnetic interaction ($J_2>0$) leads to magnetic frustration and  to a spiral spin ordering with the pitch angle $\cos \psi = J_1/4J_2$ for $J_2 > J_1/4$ at $T=0$ \cite{J1-J2}.  Without external fields, the helical order disappears for $T \neq 0$ because of the rotational SU(2) spin symmetry,  only algebraically decaying short-range order is possible.  However,  the ME interaction term, originating from a spin-orbital coupling, is loss of the inversion symmetry along the chain.  In the presence of an external electric field, the SU(2) spin symmetry is broken down to SO(2)$\times \mathbb{Z}_2$. The spin chirality, $$c_i = S_i \times S_{i+1}$$ order corresponds to the spontaneous breaking of the discrete $ \mathbb{Z}_2$ symmetry and can have a nonzero expected value  \cite{J1-J2}.  The electric dipole moments, directly relating to the spin chirality, become ordering and we have finite macroscopic FE polarization. 

\section{Results and discussion}
\begin{figure}[f]
\centering
\includegraphics[width=.7\textwidth]{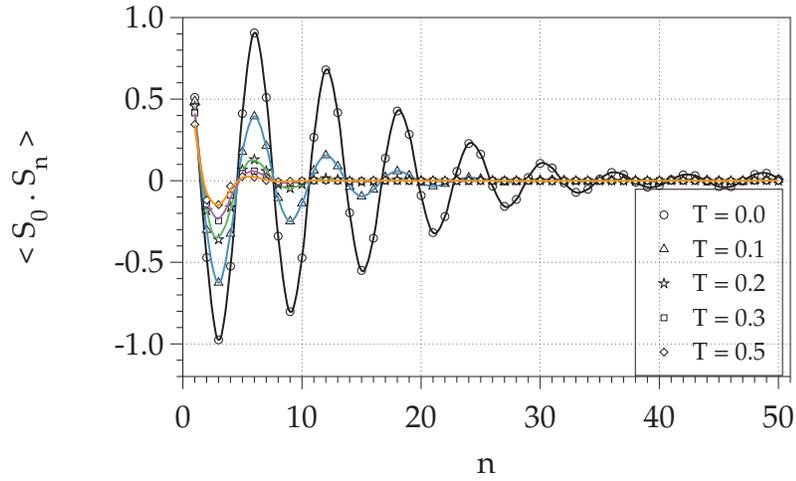}
\caption{Spin-spin correlation function as a function of the distance and the temperature in the absence of electric field. The angle between neighboring spins is $\psi = \pi /3$ given by $J_1/J_2 =2$. The solid lines are guides for the eye only.}
\label{fig:SSn}
\end{figure}

\begin{figure}[f]
\centering
\includegraphics[width=.7\textwidth]{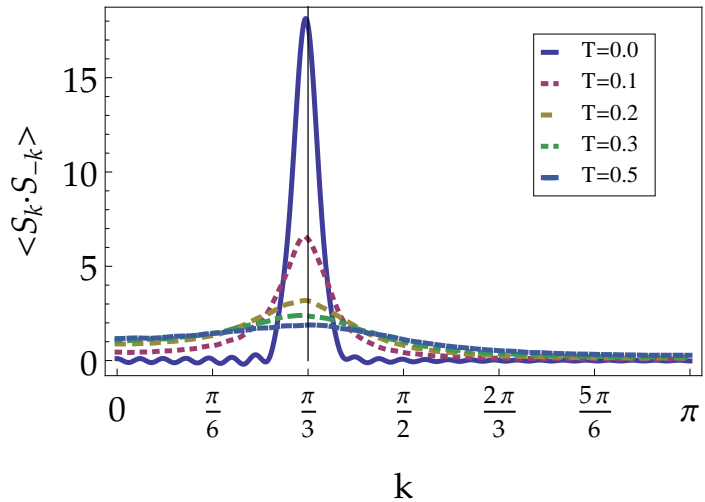}
\caption{The static spin-structure factor $\langle S_{k} \cdot S_{-k} \rangle$ for different temperatures $T/J_1$ = 0.0, 0.1, 0.2, 0.3, 0.5 without electric field $\vec E$.  The pitch angle is demonstrated by the maximum at $k = \pi/3$.}
\label{fig:SSk}
\end{figure}

\begin{figure}[f]
\centering
\includegraphics[width=.7\textwidth]{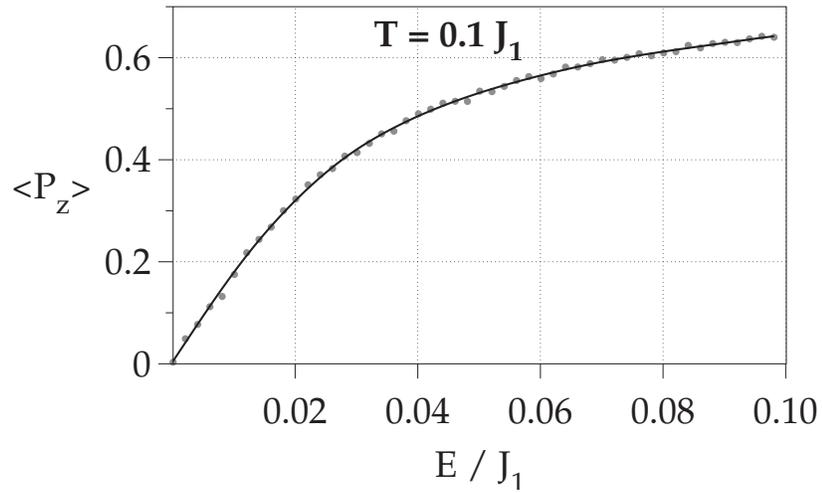}
\caption{The total FE polarization as a function of the applied electric fields E at the temperature $T =0.1 J_1$.}
\label{fig:P-E}
\end{figure}

\begin{figure}[f]
\centering
\includegraphics[width=.9\textwidth]{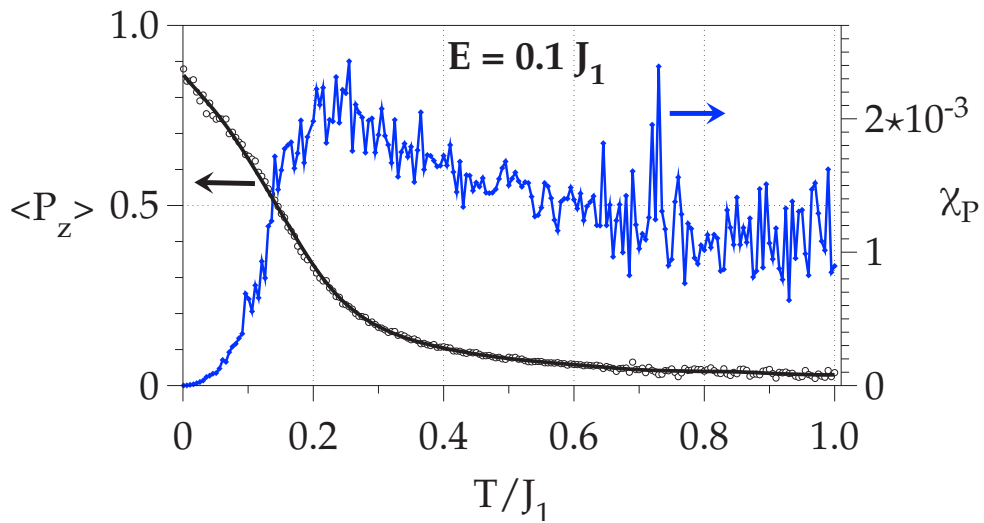}
\caption{The total FE polarization (black) and susceptibility (blue) versus temperature for the electric field $ E = 0.1 J_1$.}
\label{fig:P-T}
\end{figure}

The Monte-Carlo simulation is performed on a helimagnet chain with 250 three-dimensional classical Heisenberg spins.  Spins are randomly updated with the coherent rotation,  nucleation and Ising-like inversion for each Monte Carlo step in order to guarantee that all possible reversal mechanism may occurs in the system and can be simulated efficiently \cite{MC}.  This is especially important in dipolar and chiral system as these interactions favors large angles between neighboring spins.  Let us now look at the data for the static correlation functions $\langle S_0 \cdot S_n \rangle$, $n = 1,2, \ldots,50$ for $J_1/J_2 =2$, cf. Fig.\ref{fig:SSn}. The algebraically decaying of the helical order is clear reflected in the amplitude of the correlation function.  As expected, the correlation becomes stronger as the temperature decreases.  The quantity of  more fundamental static spin-structure factor can be obtained by the  Fourier-transformed spin correlation function:
\begin{equation}
\langle S_k \cdot S_{-k} \rangle = 1 + \sum_{n=1}^{M}\langle S_0 \cdot S_n \rangle  \cos kn 
\end{equation}
Theoretically, $M$ should be infinite but in actual calculations we have taken $M=50$.  The pitch angle is located where the static spin-structure factor is peaked.  Fig.\ref{fig:SSk} shows the maximum of $\langle S_k \cdot S_{-k} \rangle$ tends to diverge at $T=0$,  indicating the helical long-range order.  At low temperature, the ground state $k = \psi$ and a few around excited states are accessible.  As temperature increases, the contribution of the critical wave vector $k = \psi$ decreases.  The correlation length of the chirality order shrinks as well, we thus have more multiferroic domains. On the other hand, applying even a small electric field can elongate the FE order (Fig.\ref{fig:P-E}).  Unfortunately, the long-range FE order will destroyed by the thermal fluctuations too.  The transition temperature is electric field-dependence \cite{Sirker10}. In Fig.\ref{fig:P-T}, we present the result of the polarization (spin chirality) susceptibility.  For $E = 0.1 J_1$, the susceptibility shows a broad maximum at $T = 0.25 J_1$.

\section{Conclusion}
In summary, the thermodynamics of multiferroic spin chain is studied by Monte-Carlo simulation based on the classical $J_1 - J_2$ model.  We show that the low temperature excitations are dominated by the multiferroic domain wall. The ME polarizing procedure can be applied to lift the spin chiral symmetry, and then produce a single multiferroic domain.  The role of the exchange-striction,  magnetic anisotropy,  and FE dipolar coupling for the ME thermodynamics will be investigated further. 

%%%%%%%%%%%%%%%%%%%%%%%%%%%%%%%%%%%%%%%%%%%%%%
%%%%%%%%%%%%%%%%%%%%%%%%%%%%%%%%%%%%%%%%%%%%%%
%%%%%%%%%%%%%%%%%%%%%%%%%%%%%%%%%%%%%%%%%%%%%%
%%%%%%%%%%%%%%%%%%%%%%%%%%%%%%%%%%%%%%%%%%%%%%

%%%%%%%%%%%%%%%%%%%%%%%%%%%%%%%%%%%%%%%%%%%%%%
%%%%%%%%%%%%%%%%%%%%%%%%%%%%%%%%%%%%%%%%%%%%%%

%%%%%%%%%%%%%%%%%%%%%%%%%%%%%%%%%%%%%%%%%%%%%%
%%%%%%%%%%%%%%%%%%%%%%%%%%%%%%%%%%%%%%%%%%%%%%
%

\end{document}